\documentclass[prl,twocolumn]{revtex4-1}%
\usepackage{amsmath}
\usepackage{graphicx}
\usepackage[latin9]{inputenc}
\usepackage[T1]{fontenc}
\usepackage[english]{babel}
\usepackage{amsfonts}
\usepackage{amssymb}
\usepackage{color}%
\providecommand{\U}[1]{\protect\rule{.1in}{.1in}}
\newcommand{\be}{\begin{equation}}
\newcommand{\ee}{\end{equation}}
\newcommand{\ben}{\begin{eqnarray}}
\newcommand{\een}{\end{eqnarray}}

\ifx\pdfoutput\relax\let\pdfoutput=\undefined\fi
\newcount\msipdfoutput
\ifx\pdfoutput\undefined\else
\ifcase\pdfoutput\else
\ifx\paperwidth\undefined\else
\ifdim\paperheight=0pt\relax\else\pdfpageheight\paperheight\fi
\ifdim\paperwidth=0pt\relax\else\pdfpagewidth\paperwidth\fi
\fi\fi\fi
\begin{document}
\title{Imaginary-time nonuniform mesh method for solving the multidimensional
Schr\"odinger equation: Fermionization and melting of quantum Lennard-Jones crystals}
\author{Alberto Hernando}
\email{alberto.hernandodecastro@epfl.ch}
\author{Ji\v{r}\' i Van\'i\v{c}ek}
\email{jiri.vanicek@epfl.ch}
\affiliation{Laboratory of Theoretical Physical Chemistry, Institut des Sciences et
Ing\'enierie Chimiques, \'Ecole Polytechnique F\'ed\'erale de Lausanne,
CH-1015 Lausanne, Switzerland}

\begin{abstract}
An imaginary-time nonuniform mesh method is presented and used to find the
first 50 eigenstates and energies of up to five strongly interacting spinless
quantum Lennard-Jones particles trapped in a one-dimensional harmonic
potential. We show that the use of tailored grids reduces drastically the
computational effort needed to diagonalize the Hamiltonian and results in a
favorable scaling with dimensionality. Solutions to both bosonic and fermionic
counterparts of this strongly interacting system are obtained, the bosonic
case clustering as a Tonks-Girardeau crystal exhibiting the phenomenon of
fermionization. The numerically exact excited states are used to describe the
melting of this crystal at finite temperature.

\end{abstract}
\date{\today}
\maketitle

The multidimensional Schr\"odinger equation (MDSE) is undoubtedly one of the
cornerstones of modern physics and much attention has been paid to developing
efficient numerical methods for finding its solutions
\cite{Harris:1965,Kosloff:1988,Shimshovitz:2012,Pollet:2012,book_Parr_Yang,
Dalfovo:1995,Savenko:2013,Rogel-Salazar:2013,Billing:2000,Runge_Gross:1984,Abid:2003,book_Wyatt,
Alvarez:2011,Tkatchenko:2012,Delgado:1996,Corney:2004,Torres-Vega:1991,McMahon:2012}%
. A very rich testing ground for such methods has been provided by the
observation of new quantum phases at ultracold temperatures in finite and
homogeneous systems
\cite{McMahon:2012,Kapitza:1938,Leggett:1972,Grebenev:2000,Anglin_Ketterle:2002,Balibar:2010}%
, and also by the development of optical lattices where ultracold atoms are
trapped \cite{Bloch:2005}. Due to their fascinating structural and dynamical
properties, special attention has been recently devoted to one-dimensional
traps
\cite{Ronzheimer:2013,Panfil:2013,Vignolo:2013,Gring:2012,Kinoshita:2006,Paredes:2004}%
. Indeed, in the strongly interacting (Tonks-Girardeau) regime of bosonic
particles trapped in one-dimensional geometries, the repulsive nature of the
atomic interaction at short distances gives rise to the phenomenon known as
\textit{fermionization}, the mechanism of which is actively studied both
theoretically and experimentally. 

Rigorous description and explanation of the new physics found in these
well-controlled experiments require accurate theoretical methods and
constitute a formidable challenge \cite{Bloch:2008}, the main technical
difficulty being the scaling of numerical algorithms with the number of
dimensions $D$. Indeed, standard algorithms for solving differential
equations, such as the Finite Difference method, scale exponentially with
dimensions \cite{book_Morton_Mayers}, making numerical solutions of
many-dimensional problems impracticable, if not impossible. Improved methods
addressing this difficulty in the case of stationary states include the
Discrete Variable Representation (DVR) \cite{Harris:1965}, collocation method
\cite{Kosloff:1988}, phase-space method based on von Neumann periodic lattice
\cite{Shimshovitz:2012}, variational or diffusion quantum Monte Carlo (MC)
methods \cite{McMahon:2012,Pollet:2012}, Density Functional Theory (DFT)
\cite{McMahon:2012,book_Parr_Yang}, mean-field or pseudopotential interaction
models \cite{Dalfovo:1995,Savenko:2013,Rogel-Salazar:2013}, and many others.
Some of these methods find only the ground state of the time-independent MDSE,
using different efficient techniques such as the imaginary time (IT)
propagation \cite{Popov:2005} or the Variational Principle \cite{book_Sakurai}%
. Methods for real-time quantum dynamics include the Time-Dependent DVR
\cite{Billing:2000}, DFT \cite{Runge_Gross:1984}, mean-field approaches
\cite{Abid:2003}, trajectory-based methods such as Bohmian dynamics
\cite{book_Wyatt}, or time-dependent density matrix renormalization group
(t-DMRG) method \cite{Alvarez:2011}, which has proven to be very efficient in
one-dimensional geometries. Despite many accomplishments in special cases,
finding excited states and describing the real-time dynamics governed by a
general high-dimensional Hamiltonian in the strongly interacting regime
remains a difficult computational challenge.

In this paper \textsl{we propose a novel general method, scaling favorably
with dimensions, which is able to solve the time-independent MDSE numerically
exactly and simultaneously finds both its ground and excited states}.
Obviously, the proposed IT nonuniform mesh method (ITNUMM) is not intended to
replace other well established approaches; instead we expect it to have a
domain of applicability where other methods present more technical
difficulties, such as in finding excited states of many-dimensional systems
and where efficiency is more important than high accuracy. To show that ITNUMM
achieves these goals, we apply it to find the wavefunctions of the first 50
states of an ensemble of up to five distinguishable Lennard-Jones (LJ)
spinless particles trapped in a one-dimensional harmonic potential in the
Tonks-Girardeau regime. Once these states are obtained, we find, via
symmetrization and anti-symmetrization, the solutions for the Bose-Einstein
and Fermi-Dirac statistics, respectively, and observe fermionization in the
bosonic case. We also show that the computed excited states can be used in a
thermal average to describe the melting of the LJ\ clusters at finite
temperature. \textsl{As we use no other approximation than the numerical
discretization of space and time, the obtained results are numerically exact}.

The derivation of our method starts by rewriting the time-dependent MDSE
\cite{book_Sakurai}
\begin{equation}
i\hbar\frac{d}{dt}|\psi(t)\rangle=\mathcal{H}|\psi(t)\rangle,
\end{equation}
where $|\psi(t)\rangle$ is the quantum state at time $t$ of the $D$%
-dimensional system described by Hamiltonian $\mathcal{H}$, in terms of the
quantum propagator $K({\mathbf{q}},{\mathbf{q}}^{\prime};t-t^{\prime
}):=\langle{\mathbf{q}}|e^{-i(t-t^{\prime})\mathcal{H}/\hbar}|{\mathbf{q}%
}^{\prime}\rangle$ in the position basis $|{\mathbf{q}}\rangle$:%
\begin{equation}
\psi({\mathbf{q}},t)=\int d{\mathbf{q}}^{\prime}K({\mathbf{q}},{\mathbf{q}%
}^{\prime};t-t^{\prime})\psi({\mathbf{q}}^{\prime},t^{\prime}).
\end{equation}
Hamiltonian $\mathcal{H}:=\mathcal{H}_{0}+\mathcal{H}_{1}$ is now split into
two components: $\mathcal{H}_{0}$ is any Hamiltonian that includes the kinetic
energy operator $\mathcal{T}$ and whose matrix elements in the ${\mathbf{q}}%
$-representation are known, while $\mathcal{H}_{1}\equiv\mathcal{H}%
_{1}({\mathbf{q}})$ is any many-body potential depending only on ${\mathbf{q}%
}$. For very short time intervals $t-t^{\prime}=\Delta t$, the time evolution
operator can be split to first order as $e^{-i\Delta t\mathcal{H}/\hbar
}=e^{-i\Delta t\mathcal{H}_{0}/\hbar}e^{-i\Delta t\mathcal{H}_{1}/\hbar
}+O(\Delta t^{2})$ and one can write
\begin{align}
\psi({\mathbf{q}},t^{\prime}+\Delta t)  &  = \int d{\mathbf{q}}^{\prime}%
K_{0}({\mathbf{q}},{\mathbf{q}}^{\prime};\Delta t)e^{-i\Delta t\mathcal{H}%
_{1}({\mathbf{q}}^{\prime})/\hbar}\psi({\mathbf{q}}^{\prime},t^{\prime
})\nonumber\\
&  \ +O(\Delta t^{2}), \label{eq:1}%
\end{align}
where $K_{0}({\mathbf{q}},{\mathbf{q}}^{\prime};\Delta t):=\langle{\mathbf{q}%
}|e^{-i\Delta t\mathcal{H}_{0}/\hbar}|{\mathbf{q}}^{\prime}\rangle$ is the
propagator of $\mathcal{H}_{0}$, which is assumed to be known explicitly.

The $|{\mathbf{q}}\rangle$ basis is discretized as
\begin{equation}
\int d{\mathbf{q}}|{\mathbf{q}}\rangle\langle{\mathbf{q}}|=\lim_{N\rightarrow
\infty}\sum_{j=1}^{N}w({\mathbf{q}}_{j})|{\mathbf{q}}_{j}\rangle
\langle{\mathbf{q}}_{j}|
\end{equation}
where $w({\mathbf{q}}_{j})$ is a weight function depending on a particular
realization of the $N$ states $|{\mathbf{q}}_{j}\rangle$. Indeed, $w$ is
defined as $w({\mathbf{q}}):=[Np({\mathbf{q}})]^{-1}$, where $p({\mathbf{q}})$
is the density distribution of the ${\mathbf{q}}_{j}$. With this
discretization, Eq. (\ref{eq:1}) becomes
\begin{align}
&  \!\!\!\!\psi({\mathbf{q}}_{j},t^{\prime}+\Delta t)=\label{eq:2}\\
&  \lim_{N\rightarrow\infty}\sum_{k=1}^{N}w({\mathbf{q}}_{k})K_{0}%
({\mathbf{q}}_{j},{\mathbf{q}}_{k};\Delta t)e^{-i\Delta t\mathcal{H}%
_{1}({\mathbf{q}}_{k})/\hbar}\psi({\mathbf{q}}_{k},t^{\prime}).\nonumber
\end{align}
Since our main interest is finding the stationary states of $\mathcal{H}$, in
the following we will assume that (i) $\psi({\mathbf{q}},t)=e^{-itE_{n}/\hbar
}\varphi_{n}({\mathbf{q}})$ where $\varphi_{n}({\mathbf{q}})$ and $E_{n}$ are
the \mbox{$n$th} eigenstate and eigenenergy of the Hamiltonian $\mathcal{H}$,
and that (ii) the evolution is performed in IT ($t\rightarrow-i\tau$).
Although the density $p({\mathbf{q}})$ is arbitrary, below we show that Eq.
(\ref{eq:2}) simplifies in the IT scheme if this density corresponds to the
classical Boltzmann distribution of $\mathcal{H}_{1}$, namely if
\begin{equation}
p({\mathbf{q}})=Z_{\mathcal{H}_{1}}^{-1}e^{-\Delta\tau\mathcal{H}%
_{1}({\mathbf{q}})/\hbar},
\end{equation}
where $Z_{\mathcal{H}_{1}}=\mathrm{Tr}e^{-\Delta\tau\mathcal{H}_{1}/\hbar}$ is
a normalization constant (called configuration integral) and $\Delta\tau
/\hbar$ plays the role of the inverse temperature $\beta$. Under these
conditions, Eq. (\ref{eq:2}) reads
\begin{align}
&  \!\!\!\!\!\!\!\!e^{-\Delta\tau E_{n}/\hbar}\varphi_{n}({\mathbf{q}}%
_{j})=\label{eq:3}\\
&  \lim_{N\rightarrow\infty}\frac{Z_{\mathcal{H}_{1}}}{N}\sum_{k=1}^{N}%
K_{0}({\mathbf{q}}_{j},{\mathbf{q}}_{k};-i\Delta\tau)\varphi_{n}({\mathbf{q}%
}_{k}).\nonumber
\end{align}
By defining vector $\Phi_{n}:=\{\varphi_{n}({\mathbf{q}}_{j})\}_{j=1}^{N}$,
whose $j$th component is the wavefunction evaluated at position ${\mathbf{q}%
}_{j}$, and matrix $\hat{K}_{jk}:=K_{0}({\mathbf{q}}_{j},{\mathbf{q}}%
_{k};-i\Delta\tau)Z_{\mathcal{H}_{1}}/N$ whose elements are proportional to
the propagator $K_{0}$ from ${\mathbf{q}}_{j}$ to ${\mathbf{q}}_{k}$, one can
rewrite Eq. (\ref{eq:3}) as a matrix eigenvalue equation
\begin{equation}
e^{-\Delta\tau E_{n}/\hbar}\Phi_{n}=\hat{K}\cdot\Phi_{n}.
\end{equation}
This equation, central to the ITNUMM, exhibits the main advantage of our
method---\textsl{the problem of finding the spectrum and eigenfunctions of the
original Hamiltonian $\mathcal{H}$ is reduced to sampling the classical
Boltzmann distribution and diagonalizing $\hat{K}$ evaluated at those points}.
Instead of the Hamiltonian, we diagonalize the imaginary-time propagator,
i.e., a matrix with analytically known and real-valued elements.
\textsl{Evaluation of, e.g., derivatives or Fourier transforms is not needed}.
Indeed, the implementation of the algorithm is rather simple since it only
requires standard methods for sampling from arbitrary probability
distributions and diagonalizing sparse real-valued matrices. The computational
effort is also reduced by constructing a nonuniform grid in which more grid
points are placed in areas where the wavefunctions exhibit more detailed
features. In the special case of $\mathcal{H}_{0}\equiv\mathcal{T}$,
$\mathcal{H}_{1}({\mathbf{q}})$ equals the classical potential energy, $K_{0}$
is a free-particle propagator in $D$ dimensions \cite{book_Sakurai}, and
matrix elements $\hat{K}_{jk}$ assume the Gaussian form
\begin{equation}
\hat{K}_{jk}=\frac{Z_{\mathcal{H}_{1}}}{N}\left(  \frac{m}{2\pi\hbar\Delta
\tau}\right)  ^{D/2}\exp\left[  -\frac{m}{2\hbar\Delta\tau}({\mathbf{q}}%
_{j}-{\mathbf{q}}_{k})^{2}\right]  ,
\end{equation}
where $m$ is the mass, for simplicity assumed to be the same for all degrees
of freedom. In correlated systems, where sampling the Boltzmann distribution
is difficult or unfeasible---as in the case of Coulomb interaction, we propose
the splitting $\mathcal{H}_{1}({\mathbf{q}})=V_{1}({\mathbf{q}})+V_{2}%
({\mathbf{q}})$, where $V_{1}({\mathbf{q}})$ is a sum of well-behaved one-body
potentials and $V_{2}({\mathbf{q}})$ is the remainder including all
correlations. Here the sampling is performed with the weight $p({\mathbf{q}%
})=Z_{V_{1}}^{-1}e^{-\Delta\tau V_{1}({\mathbf{q}})/\hbar}$ and normalization
$Z_{V_{1}}=\mathrm{Tr}e^{-\Delta\tau V_{1}/\hbar}$; the matrix to be
diagonalized becomes
\begin{align}
\hat{K}_{jk}  &  =\frac{Z_{V_{1}}}{N}\left(  \frac{m}{2\pi\hbar\Delta\tau
}\right)  ^{D/2}\nonumber\\
&  \times\exp\left[  -\frac{m}{2\hbar\Delta\tau}({\mathbf{q}}_{j}-{\mathbf{q}%
}_{k})^{2}-\frac{\Delta\tau}{\hbar}V_{2}({\mathbf{q}}_{k})\right]  .
\label{eq:4}%
\end{align}

We have found this method to be very efficient in one-dimensional problems
with several very different potentials. Although an arbitrary sampling
procedure can be used, we have employed a \textit{quadrature scheme}: instead
of random sampling of $p(q)$ by a MC procedure, the $q_{j}$ points are chosen
with a deterministic algorithm. The motivation for this approach is reducing
to a minimum the number of vector-elements needed for a given accuracy, and
thus reducing the computational cost of the diagonalization of $\hat{K}$.
Specifically, we first consider a new variable $u$, uniformly distributed in
the interval $[0,1]$, and define an equidistant grid $u_{j}=(j-1/2)/N$. The
Jacobian of the transformation from $q$ to $u$ is given by $p(q)$ since
$p(q)dq=du$, hence
\begin{equation}
u(q)=\int_{-\infty}^{q}dq^{\prime}p(q^{\prime})=P(q),
\end{equation}
where $P(q)$ is the cumulative distribution function. Next, the $q$-grid is
obtained by inverting this equation for all values of $u_{j}$, and once the
$q$-grid is ready, the evaluation and diagonalization of the matrix $\hat{K}$
is performed with standard numerical methods.

As the first application of ITNUMM, we solved (i) the 1D harmonic oscillator
\cite{book_Sakurai} $\mathcal{H}_{1}(q)=m\omega^{2}q^{2}/2$, using natural
units for energy and position (defined by $\hbar\omega$ and $\sqrt
{\hbar/m\omega}$, respectively), and (ii) two particles of equal mass $m$
interacting via a LJ potential $\mathcal{H}_{1}(q)=V_{\mathrm{LJ}}%
(q)\equiv\epsilon\lbrack(r_{e}/q)^{12}-2(r_{e}/q)^{6}]$. For the latter, we
used a de Boer quantum delocalization length \cite{Deckman:2009} of
$\Lambda=2^{1/6}\hbar/(r_{e}\sqrt{m\epsilon})=0.16$, corresponding to
hypothetical particles with properties between para-hydrogen ---where quantum
effects dominate---and neon---where quantum effects are present but classical
behavior dominates. In the Supplementary Material (SM), we show the grid
points, eigenvalues, and several eigenstates obtained with ITNUMM in both
cases---we also include a notebook executable in the Wolfram Research's
Mathematica software, where the interested reader can explore the technical
details of the method.

\begin{figure}
[tb]\includegraphics[width=\hsize,trim= 32 27 20 20,clip=true]{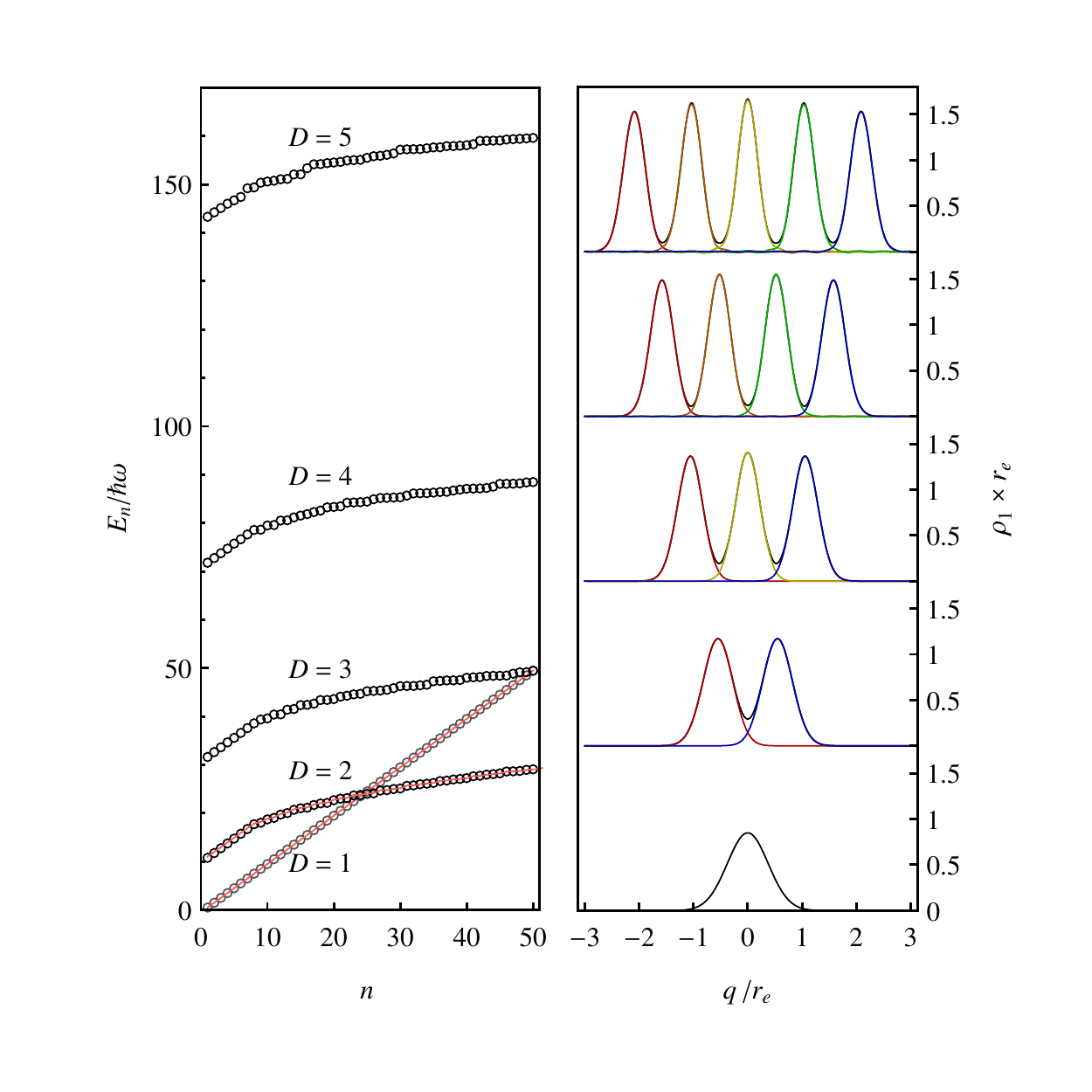}

\caption{ (Color online) Left: Energy spectrum for $D$ LJ particles in a 1D
harmonic trap obtained with our method (circles). The exact results for $D=1$
and $2$ are shown as red solid lines. Energies are shifted to the minimum
of the potential $\min(\mathcal{H}_1)=-\protect\epsilon(D-1)D/2$. Right:
One-body densities (normalized to the number of particles) of the ground state for
distinguishable (colored lines) and indistinguishable particles (black
lines).} \label{fig:eneden}
\end{figure}

As expected, we observed that the imaginary time $\Delta\tau$ must be small
enough to reduce the relative error $\sigma$ introduced by the splitting of
the propagator---which is $\sigma\sim O(\Delta\tau^{3})$ since the second
order term vanishes for stationary states---but large enough to avoid reducing
the Gaussian elements of the $\hat{K}$ matrix to delta functions and
eventually obtaining a diagonal matrix. The latter condition is ensured by
requiring $1\gg\hat{K}_{jj}=Z_{\mathcal{H}_{1}}\left(  m/2\pi\hbar\Delta
\tau\right)  ^{D/2}/N$, which imposes a lower bound on $\Delta\tau$ for a
given $N$. In the SM, we explore the dependence of the relative error $\sigma$
on $\Delta\tau$ for a given number $N$ of grid points, and also the dependence
of $\sigma$ on $N$ in the harmonic oscillator. Remarkably, the relative error
can be fitted to $\sigma(N)\simeq0.18N^{-1.9}$, indicating a significantly
faster convergence rate than the rate expected for a MC scheme [$\sigma(N)\sim
N^{-1/2}$] \cite{McMahon:2012,Pollet:2012}. Regarding the excited states, we
found that the error becomes large for states with the highest eigenenergies.
Indeed, the number of grid points $N$ becomes insufficient to reproduce the
characteristic high frequency oscillations of wavefunctions describing highly
excited states. Yet, the agreement with exact results is very good for the
first $150$ states using $N=500$ grid points, as shown in
Fig.~\ref{fig:eneden} for the first 50 states (the whole spectrum is shown in
the SM).

%%%%%%%%%%%%%%%%%%%%%%%%%%%%%%%%%%%%%%%%%

%%%%%%%%%%%%%%%%%%%%%%%%%%%%%%%%%%%%%%%%%

\begin{figure}
[tb]\includegraphics[width=0.9\hsize]{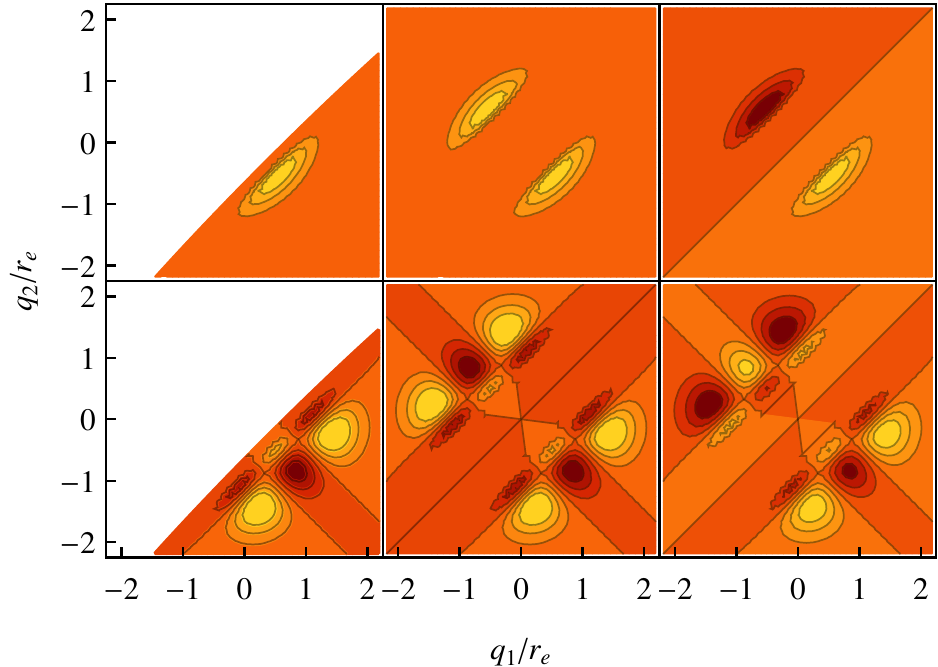}

\caption{ (Color online) Wavefunctions $\protect\varphi_n(q_1,q_2)$ of the ground and 19th states  for
$D=2$ LJ particles in a 1D harmonic trap (see text for details).
Lighter (darker) color indicates positive (negative) values of the wavefunction.
Left: distinguishable particles in the subspace $q_1>q_2$; center:
indistinguishable bosons; right: indistinguishable fermions. }
\label{fig:states2D}
\end{figure}

As a more stringent test, we now apply the method to $D$ LJ particles in a
one-dimensional harmonic trap. Potentials $V_{1}$ and $V_{2}$ are defined by
\begin{align}
V_{1}({\mathbf{q}})  &  =\sum_{\lambda=1}^{D}\frac{1}{2}m\omega^{2}q_{\lambda
}^{2},\\
V_{2}({\mathbf{q}})  &  =\sum_{\lambda<\mu}^{D}V_{\mathrm{LJ}}(|q_{\lambda
}-q_{\mu}|),
\end{align}
the de Boer length has the same value as in the example above, and $\omega
r_{e}\sqrt{m/\epsilon}=1/2$. The problem is separable only for $D=1$ or $2$,
and so a multidimensional numerical method is mandatory for $D\geq3$. In order
to reduce the number of grid points in the numerical calculation, we first
solve the problem for distinguishable particles and construct \textit{a
posteriori} the eigenstates of indistinguishable particles by symmetrizing or
anti-symmetrizing the wavefunction for spinless bosons or fermions,
respectively. Thanks to the repulsive nature of the LJ potential at short
distances we only need to evaluate $\hat{K}$ in the subspace defined by
$q_{1}>q_{2}+a,\dots,$ $q_{D-1}>q_{D}+a$, where $a$ is the core radius of the
LJ\ potential, within which the wavefunction is expected to be zero within
numerical accuracy ($a=0.63r_{e}$ in our calculations). The grid points are
sampled from the classical Boltzmann distribution of the harmonic trap in this
subspace, $p({\mathbf{q}})=Z_{V_{1}}^{-1}e^{-\Delta\tau m\omega^{2}%
|{\mathbf{q}}|^{2}/2\hbar}$ with $Z_{V_{1}}=(2\pi/\Delta\tau m\omega
^{2})^{D/2}/C_{D}(a)$, where the normalization constant obeys $C_{D}(0)=D!$.
All the two-body interactions, contained in $V_{2}({\mathbf{q}})$, are
evaluated in the matrix elements of $\hat{K}$. As mentioned above, only the
low-lying eigenstates are accurate, so we have used the Arnoldi algorithm
\cite{Arnoldi:1951} to obtain the first 50 eigenstates. We have taken into
account that many of the matrix elements are close to zero by using standard
computational techniques for sparse matrices: instead of storing the $N\times
N$ values of the matrix, only elements larger than a certain threshold were
stored. Parameters used in calculations with varying $D$ were
\begin{equation}%
\begin{array}
[c]{r|ccccc}%
D & 1 & 2 & 3 & 4 & 5\\\hline
N & 500 & 6709 & 14~394 & 36~517 & 84~690\\
\Delta\tau & 0.0055 & 0.15 & 1.5 & 1.5 & 1.5
\end{array}
\nonumber
\end{equation}
Note the relatively low total number of grid points needed to obtain results
with reasonable accuracy (a relative error of 0.002 for the $D=2$ case).
Figure~\ref{fig:states2D} shows the ground and 19th states for $D=2$ and for
the three statistics: distinguishable particles (in the above mentioned
subspace), bosons, and fermions (in the full space). The spectrum of
$\mathcal{H}$ as a function of $D$ is shown in Fig.~\ref{fig:eneden} (left
panel). We find the same spectrum for the three cases, which is a consequence
of the fermionization \cite{Paredes:2004} mechanism due to the repulsive
behavior of the LJ potential at short distances. Indeed, the bosonic and
fermionic systems show the same one-body densities in position space, as shown
in the right panel of Fig.~\ref{fig:eneden}. In all three cases the densities
show a well-defined structure, forming a quantum crystal. The displayed
one-body densities, defined as \cite{book_Lipparini}
\begin{equation}
\rho_{n}(q_{\lambda})=\int|\varphi_{n}({\mathbf{q}})|^{2}\prod_{\mu\neq
\lambda}^{D}dq_{\mu},
\end{equation}
were obtained from the nonuniform mesh as follows: first,we computed its
Fourier transform in a regular equidistant grid in momentum ($k$) space as
\begin{align}
\tilde{\rho}_{n}(k)  &  =\int|\varphi_{n}({\mathbf{q}})|^{2}e^{-ikq_{\lambda}%
}\prod_{\mu=1}^{D}dq_{\mu}\\
&  \approx\frac{Z}{N}\sum_{j=1}^{N}e^{\Delta\tau V_{1}({\mathbf{q}}_{j}%
)/\hbar-ikq_{\lambda}}|\varphi_{n}({\mathbf{q}}_{j})|^{2},
\end{align}
and then Fourier-transformed $\tilde{\rho}_{n}(k)$ back to $q_{\lambda}$-space
using standard numerical methods.

%%%%%%%%%%%%%%%%%%%%%%%%%%%%%%%%%%%%%%%%%

%%%%%%%%%%%%%%%%%%%%%%%%%%%%%%%%%%%%%%%%%

\begin{figure}
[tb]\includegraphics[width=0.9\hsize]{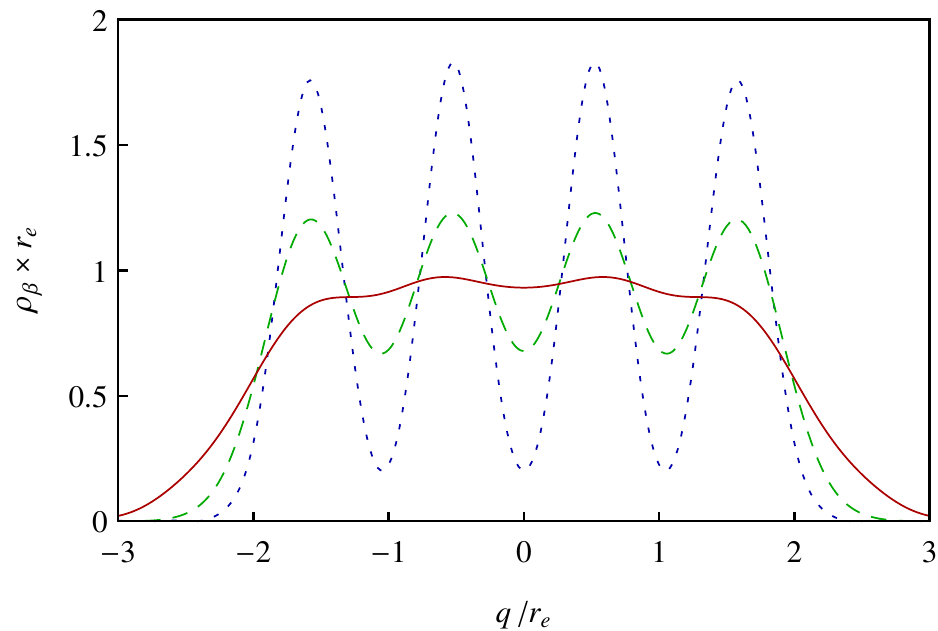}

\caption{ (Color online) One-body densities (normalized to the number of particles) for $D=4$
LJ particles in a 1D harmonic trap at three different temperatures: $\hbar
\protect\omega \protect\beta =2.8$ (dotted line), $0.7$ (dashed line), and $0.4$ (solid line).
The crystal structure disappears with increasing temperature, resulting in an
unstructured total density as in a fluid. } \label{fig:temp}
\end{figure}

The 50 states obtained in the course of the diagonalization are sufficient to
study the behavior of the system at finite temperatures. The (unnormalized)
probability distribution of the system $p_{\beta}({\mathbf{q}})$ at finite
inverse temperature $\beta$ is defined as the thermal average
\begin{equation}
p_{\beta}({\mathbf{q}})=\sum_{n=1}^{\infty}e^{-\beta E_{n}}|\varphi
_{n}({\mathbf{q}})|^{2},
\end{equation}
and the corresponding one-body density $\rho_{\beta}(q)$ is obtained similarly
as for pure states. Figure~\ref{fig:temp} shows the one-body density for $D=4$
at three different temperatures, where the lack of structure at the highest
temperature can be understood as the melting of the quantum crystal.

%%%%%%%%%%%%%%%%%%%%%%%%%%%%%%%%%%%%%%%%%

%%%%%%%%%%%%%%%%%%%%%%%%%%%%%%%%%%%%%%%%%

To summarize, we have presented compelling evidence that the proposed method
achieves the original goals. Indeed, (i) the only approximation used is the
numerical discretization of space and time; (ii) the ITNUMM only requires
standard methods for sampling from an arbitrary probability distribution and
for diagonalizing real-valued sparse matrices; (iii) both ground and excited
states are obtained in the course of the diagonalization; and (iv) due to the
nonuniform nature of the grid that uses the potential to guide the sampling,
the complexity of the algorithm is significantly reduced in high-dimensional
systems. In particular, all our calculations were performed on a single
workstation with a 64-bit 2.4 GHz Quad-Core Intel Xeon E5 processor and 12 GB
of memory. Yet, the algorithm can be easily accelerated by parallelization.
The accuracy of ITNUMM can be increased by using tailored grids, larger $N$
values, or splitting methods of a higher order than in Eq. (\ref{eq:1}). In
addition to computing thermal averages---as shown here---the large set of
excited states can be also used for solving real-time quantum dynamics in a
straightforward fashion. As we have not found any \textit{a priori} limitation
to the applicability of the method, other systems described by the MDSE will
be studied in the future.

\textit{Acknowledgments.} The authors thank E. Zambrano, M. Wehrle, M. \v{S}ulc,
and F. Mazzanti for discussions. This research was supported by the Swiss NSF
NCCR MUST (Molecular Ultrafast Science \& Technology) and by the EPFL.

\bibliographystyle{apsrev4-1}
\bibliography{method_Alb}

\end{document}